\documentstyle[twoside,fleqn,npb,amssymb,epsfig]{article}
\title{GRSV parton densities revisited}

\author{M.\ Stratmann\address{Department of Physics, University of Durham,
                              Durham DH1 3LE, England}%
        \thanks{I acknowledge travel support from the EU Forth
         Framework Programme `Training and Mobility of Researchers',
         Network `QCD and the Deep Structure of Elementary Particles',
         contract FMRX-CT98-0194 (DG 12 - MIHT).}}

\begin{document}

\begin{abstract}
An updated next-to-leading order (NLO) QCD analysis of all presently 
available longitudinally polarized deep-inelastic scattering (DIS) data
is presented in the framework of the radiative parton model.
\end{abstract}

\maketitle

%%%%%%%%%%%%%%%%%%%%%%
\section{INTRODUCTION}
%%%%%%%%%%%%%%%%%%%%%%

Many new or improved measurements of the spin asymmetry
\begin{equation}
\label{eq:a1}
A_1(x,Q^2)\simeq \frac{g_1(x,Q^2)}{F_2(x,Q^2)/[2x(1+R(x,Q^2))]},
\end{equation}
$R=F_L/2xF_1$, in longitudinally polarized DIS off various
different targets $(p,\,n,\,d)$ have become available in the past two
years: E142, E143, and SMC have presented their final data sets
\cite{ref:datafinal}, and new results from HERMES, 
E154, and E155 have been reported recently \cite{ref:datanew}.
Since none of these measurements were included in our original NLO QCD
analysis \cite{ref:grsv}, it seems to be appropriate to reanalyse these
data\footnote{The E155 proton data are still preliminary and hence not 
included in our analysis.} within the framework of \cite{ref:grsv}.

In NLO the polarized structure function 
$g_1$ in (\ref{eq:a1}) reads (suppressing all $x$ and $Q^2$ dependence)
\begin{eqnarray}
\label{eq:g1}
g_1 \!\!&=& \!\!  \frac{1}{2} \sum_{q=u,d,s} e_q^2 \left[
\left(\Delta q + \Delta \bar{q}\right) \otimes 
\left(1+\frac{\alpha_s}{2\pi} \Delta C_q\right) \right. \nonumber \\ 
&&\!\!+ \left. \frac{\alpha_s}{2\pi} \Delta g \otimes \Delta C_g \right]\,\,,
\end{eqnarray} 
where $\Delta C_{q,g}$ are the spin-dependent Wilson coefficients and
the symbol $\otimes$ denotes the usual convolution in $x$ space.
From (\ref{eq:g1}) it is obvious that inclusive DIS data
\cite{ref:datafinal,ref:datanew} can reveal
only information on $\Delta q + \Delta \bar{q}$ 
but neither on $\Delta q$ {\em and} $\Delta \bar{q}$ nor
on $\Delta g$, which enters (\ref{eq:g1}) only
as an ${\cal{O}}(\alpha_s)$ correction.
Thus one can either stick to a comprehensive analysis of polarized
DIS \cite{ref:abfr,ref:smcqcd} or one has to impose certain
{\em assumptions} about the flavor decomposition in order to 
be able to estimate processes
other than DIS for upcoming experiments like RHIC.
As in \cite{ref:grsv}, and other more recent QCD analyses 
\cite{ref:dss,ref:leader}, we follow the latter option. 

%%%%%%%%%%%%%%%%%%%%%%%%%%%%%%%%%%%%%%
\section{DETAILS OF THE GRSV ANALYSIS}
%%%%%%%%%%%%%%%%%%%%%%%%%%%%%%%%%%%%%%

Due to the lack of space we shall be rather brief and concentrate
only on the most important changes since \cite{ref:grsv}.
The limited amount of data demands a reasonably simple, but flexible
enough ansatz for the polarized densities such as 
\begin{equation}
\label{eq:ansatz}
\Delta f(x,\mu^2) = N_f x^{\alpha_f} (1-x)^{\beta_f} f(x,\mu^2)\;,
\end{equation}
with $f=u,\bar{u},d,\bar{d},s,\bar{s},g$.
Note that in \cite{ref:grsv} we actually used $f=u_v,d_v$ 
in (\ref{eq:ansatz}) instead of $u$ and $d$, 
however the positivity constraint
\begin{equation}
\label{eq:pos}
\left| \Delta f(x,Q^2) \right| \le f(x,Q^2)\;,
\end{equation}
which we want to exploit in our analysis, does not necessarily hold for the 
valence densities.
Of course, the bound (\ref{eq:pos}) is strictly
valid only in LO and is subject to NLO corrections \cite{ref:afr}
because the $\Delta f$ become unphysical, scheme-dependent objects in NLO.
However the corrections are not very pronounced, in particular at
large $x$ \cite{ref:afr}, the only region where (\ref{eq:pos}) imposes some
restrictions in practice. We therefore use (\ref{eq:pos}) also in NLO.

To further simplify (\ref{eq:ansatz}) we {\em assume} that 
$\Delta \bar{q}=\Delta \bar{u}=\Delta \bar{d}$, 
fix $N_{u,d}$ by the relations between
the first moments of the non-singlet combinations $\Delta q_{3,8}$ and 
the $F$ and $D$ values (using the updated value for 
$|g_A/g_V| = 1.2670\pm0.0035$ \cite{ref:pdg}), and take
$\Delta s = \Delta \bar{s} = \lambda \Delta \bar{q}$.
In the latter relation we {\em choose} $\lambda=1$ ($SU(3)_f$ 
symmetric sea), but similarly
agreeable fits are obtained, e.g., for $\lambda=1/2$ 
(see also \cite{ref:leader}), as well as by using an independent 
$x$ shape for $\Delta s$, reflecting the above mentioned
uncertainty in the flavor separation.

For the unpolarized distributions $f$ in (\ref{eq:ansatz}) and
(\ref{eq:pos}) we use the updated GRV densities \cite{ref:grv98}
and also adopt their values for the input scale $\mu$ 
($\approx 0.6\,\mathrm{GeV}$ in
NLO) and $\alpha_s(M_z^2)=0.114$. Note that in \cite{ref:grv98}
the RG equation for $\alpha_s$ is now solved exactly instead of using the
approximative NLO formula (see, e.g., \cite{ref:pdg}),
which is more appropriate at low $Q^2$ where many of the polarized data lie.

To fix the remaining parameters in (\ref{eq:ansatz}) we perform fits
to the directly measured spin asymmetry (\ref{eq:a1}) in 
LO {\em{and}} NLO, which is mandatory if one wants
to adopt these distributions in a consistent analysis of the
perturbative stability of polarized processes. Also most MC programs
only contain LO matrix elements, and hence LO densities are
more appropriate.
Apart from the above outlined `standard scenario' fit, equally good fits
can be performed in a `valence scenario' (see \cite{ref:grsv}) which is
based on the assumption \cite{ref:lipkin} that the $F,D$ values fix 
only the valence parts of $\Delta q_{3,8}$. The main feature
of such a model is the possibility to describe the data with a
vanishing strange sea. Due to the lack of space we do not pursue this 
scenario here and restrict ourselves in what follows to the results obtained 
in the NLO $(\overline{\mathrm{MS}})$ `standard scenario' framework.

%%%%%%%%%%%%%%%%%
\section{RESULTS}
%%%%%%%%%%%%%%%%%

\begin{figure}[htb]
\epsfig{file=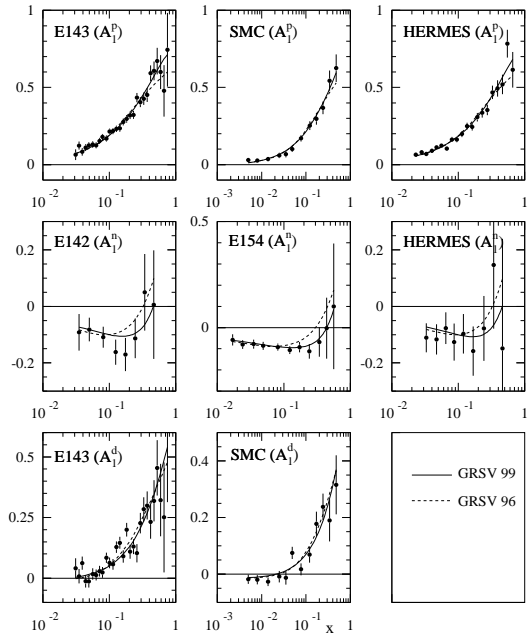,width=7.0cm}
\vspace*{-0.8cm}
\caption{Comparison of our reanalysed NLO results for
$A_1(x,Q^2)$ (solid lines) with available data
sets (not shown, but included in the fit: E155 (deuteron) and the old EMC and
SLAC data) and with our previous analysis \cite{ref:grsv} (dashed lines).}
\label{fig:fig1}
\vspace*{-0.5cm}
\end{figure}
A comparison of our new NLO fit with the available  
$A_1(x,Q^2)$ data \cite{ref:datafinal,ref:datanew} and with our
previous analysis \cite{ref:grsv} is presented in Fig.~1. The total
$\chi^2$ values of the new and old fits are 147.4 and 183, respectively,
for 185 data points and adding statistical and systematical errors 
in quadrature.
As can be seen, sizeable differences appear only in case of the 
neutron asymmetry, which leaves its footprint also in the
individual parton densities $\Delta f$ shown in Fig.~2.
Since the neutron data mainly probe $\Delta d$, the most prominent 
changes are observed here.
\begin{figure}[htb]
\vspace*{-0.5cm}
\epsfig{file=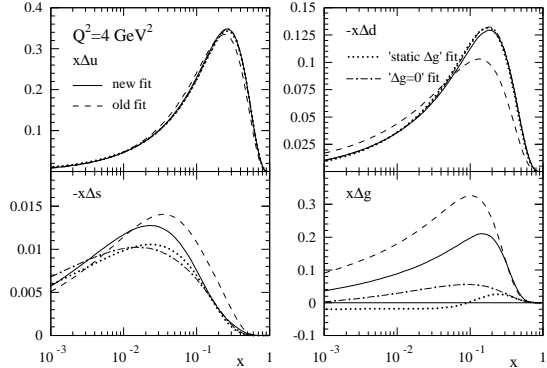,width=8.0cm}
\vspace*{-1.4cm}
\caption{The polarized NLO $\overline{\mathrm{MS}}$ densities at
$Q^2=4\,\mathrm{GeV}^2$ as obtained in the
new and old \cite{ref:grsv} GRSV analyses. Also shown are the 
distributions obtained in two other fits employing additional constraints
on $\Delta g$ (see text). In all fits shown we have chosen
a flavor symmetric sea.}
\label{fig:fig2}
\vspace*{-0.5cm}
\end{figure}

The differences in the sea and in $\Delta g$  only reflect the fact that
they are constrained to a much lesser extent by the data than $u$ and
$d$. In particular $\Delta g$ remains to be hardly constrained at all,
which is not surprising due to the lack of any direct information 
on $\Delta g$ so far.
Therefore we show in Fig.~2 also the results of two other fits, which
are based on additional constraints on $\Delta g$. For the
`$\Delta g=0$' fit we start from a vanishing gluon 
input in (\ref{eq:ansatz}), and the `static $\Delta g$' 
is chosen in such a way that its first moment becomes independent of $Q^2$. 
This can be achieved by setting $d\Delta g(Q^2)/d\ln Q^2=0$ and yields
in LO $\Delta g \simeq -\Delta \Sigma/2$ (see \cite{ref:zeuthen97}), where 
$\Delta \Sigma$ is the total helicity carried by quarks (the relation is
only subject to a small NLO correction).
Both gluons give also excellent fits to the available data and do not
affect the results for $u$ and $d$ (see Fig.~2). In fact we
can obtain fits without any significant change in $\chi^2$ for 
$\Delta g(\mu^2)$ in the range -0.3\ldots0.6 (our best fit has 0.28, 
corresponding to $\Delta g(10\,\mathrm{GeV}^2)\simeq 0.9$).
The uncertainty in $\Delta g$ is compatible to the one found 
in \cite{ref:smcqcd}, although our gluons extend also to slightly negative 
first moments. 
It is interesting to observe that for our best fit gluon the spin of
the nucleon 
\begin{equation}
\label{eq:spinsum}
S_z=\frac{1}{2} = \frac{1}{2} \Delta \Sigma(Q^2)+\Delta g(Q^2)+L_z(Q^2)
\end{equation}
is dominantly carried by quarks and gluons at our input scale $\mu$, and
only during the $Q^2$ evolution a large negative $L_z(Q^2)$ is being
built up in order to compensate for the strong rise of 
$\Delta g(Q^2)$, see Fig.~5 in \cite{ref:zeuthen97}.
For the `static $\Delta g$' the situation is completely different:
by construction the quark and gluon contributions to (\ref{eq:spinsum})
cancel each other implying that for {\em all} values of $Q^2$ the spin
is entirely of angular momentum origin, contrary to what is
intuitively expected. 

Inevitably the large uncertainty in $\Delta g$ implies that the small $x$
behaviour of $g_1$ is completely uncertain and not predictable 
(see Fig.~3 in \cite{ref:zeuthen97}), which translates also into a large 
theoretical error from the $x\rightarrow 0$ extrapolation
when calculating first moments of $g_1$. 
Taking our best fit we obtain $\Gamma_1^p=0.133$ and $\Gamma_1^n=-0.624$ 
at $Q^2=10\,\mathrm{GeV}^2$ in agreement with a recent SMC QCD 
analysis \cite{ref:smcqcd}.

The two challenging questions concerning polarized parton densities 
are still $\Delta g$ and the flavor decomposition. Recent semi-inclusive DIS
results \cite{ref:ruh} may help to unravel the latter, but major
progress, in particular on $\Delta g$, can be only expected from RHIC.
A realization of the optional upgrade of HERA to a polarized collider
would be also very helpful to gain more insight into the spin
structure of the nucleon {\em and} the photon, with the latter
being completely unmeasured so far.

\section*{ACKNOWLEDGMENTS}
It is a pleasure to thank M.\ Gl\"{u}ck, E.\ Reya, and
W.\ Vogelsang for a fruitful collaboration.

\end{document}